\documentclass{article}
\pdfoutput=1
\usepackage{jheppub}
\usepackage{lineno}
\newcommand{\sgn}{\text{sgn}}

\newcommand{\sample}{\text{sample}}
\title{A computational proof of locality in entanglement.}
\author{Han Geurdes,}
\affiliation{
Geurdes datascience, KvK64522202, C. vd Lijnstraat 164 2593 NN Den Haag, Netherlands}
\emailAdd{han.geurdes@gmail.com}
\abstract{
In this paper the design and proof of concept (POC) coding of a local hidden variables computer model is presented. The program violates the Clauser, Horne, Shimony and Holt inequality $|$CHSH$|$ $\leq 2$. In our numerical experiment, we find with our local computer program,  CHSH $\approx 1 + \sqrt{2}$.  } 
\begin{document}
\maketitle

\section{Introduction.}
\subsection{Tabula rasa}
In introductory courses of quantum mechanics, foundation of quantum theory is a difficult topic. A general presentation of quantum mechanical interpretations can be found in \cite{Jammer}. 
In a good but somewhat older textbook such as Merzbacher \cite{Merz}, the probability interpretation is flatly introduced as a doctrine.  Given the wave function $\psi(\mathbf{r},t)$, the doctrine is that the probability to find a particle in a volume $d^3r$ around a point $\mathbf{r}$ in $\mathbb{R}^3$, at time $t$, is equal to $|\psi(\mathbf{r},t)|^2d^3r$. Here, $|\psi(\mathbf{r},t)|^2=\psi^{*}(\mathbf{r},t)\psi(\mathbf{r},t)$.
The approach of Hameka follows, page 20, a similar procedure \cite{Hameka} as does the textbook, page 73, of Rae, \cite{Rae}.

It remains a mystery why nature has two different types of probability. It also is a mystery how the one type of probability transforms to the other and why addition of relativity bars the possibility of a wave interpretation of the wave function \cite{Grein}.

An even bigger mystery in quantum theory is entanglement. Perhaps that some explanation is at its place here. Entanglement is the translation of the term "Verschr{\"a}nkung" introduced by
Schr{\"o}dinger. It means that there are pure states of a compound system which
yield stronger correlations in the joint probability distribution of measuring results
on the subsystems than those which can arise from correlations between
individual states of the subsystems \cite{Haag}. 
 
\subsection{Bell's work}
In the debate of the foundation of quantum theory, Bell's theorem \cite {Bell} is considered an important milestone. In order to study Einstein’s incompleteness criticism \cite{Eins}, Bell formulated an expression for the correlation between distant spin measurements. With this formulation it was possible to answer Einstein's question of completeness with an experiment. It is important to note the following. The experimenters using Bell's correlation formula did not "look under the hood" for extra parameters. They employed classical statistics in spin measurement experiments without much physics theory about hidden variables. The key element is that Bells theorem exploits the fundamental difference between measure theoretic probability and quantum probability. 

Einsteins criticism initially did not include the spin. The reformulation of Einsteins criticism \cite{Eins} into the entanglement between spins was provided by David Bohm \cite{Bohm:1983} and \cite{Bohm2}.  For the ease of the argument, let us say that Einstein argued for extra hidden parameters to explain spin correlation. Einstein insisted that the A wing of the experiment is independent of what is done in the B wing and vice versa \cite{Peres}. Funny enough, here we also can ask a naive question. Namely, how far does Einstein think A and B should be separated for this independence to occur? 

The restriction of locality was introduced because in theory the correlation is independent of the distance between the sites of measurement. The Einsteinian locality concept, -however see also the previous naive questions-,  can be tested with the use of the Clauser, Horne, Shimony and Holt (CHSH) inequality. The inequality  is derived \cite{Clauser} from Bells correlation formula \cite{Bell}, $E(a,b)$. Bells formula reads
\begin{linenomath*}\begin{equation}\label{1}
E(a,b)=\int d\lambda \rho_{\lambda}A_{\lambda}(a)B_{\lambda}(b)
\end{equation}\end{linenomath*}
In equation (\ref{1}) the (classical) probability density of the hidden variables, $\lambda$, is $\rho_{\lambda} \geq 0$. So, $\int d \lambda \rho_{\lambda}=1$. The local effect of the $\lambda$, e.g. an array $(\lambda_1,\lambda_2)$, can be accomplished if e.g. $\lambda_1$ is assigned  and related to the $A$ wing and $\lambda_2$ to the $B$ wing of the experiment. Furthermore, the measurement functions $A_{\lambda}(a)$ and $B_{\lambda}(b)$ both project in $\{-1,1\}$ to represent binairy spin variables (e.g. up=1, down=-1 along an arbitrary z-axis). The $a$ and $b$ represent unit parameter vectors. 
Given (\ref{1}) we can study the following four term;
\begin{linenomath*}\begin{equation}\label{2}
S=E(1,1)-E(1,2)-E(2,1)-E(2,2)
\end{equation}\end{linenomath*}
The CHSH inequality $S \leq 2$ can be derived from (\ref{2}). See \cite{Clauser} and e.g. \cite{Peres}. So for an $E(a,b)$ in the form (\ref{1}) we have by necessity $S \leq 2$. 
However, note that $S>2$, is possible with $E(a,b)=a \cdot b$ for certain proper $(a,b)$ combinations of setting parameter vectors. 
To be sure, the labels $1$ and $2$ in (\ref{2}) refer to $a$ and $b$ vectors that can be set in the experiment. E.g. $1$ on the A side, operated by Alice, is $a^{(1)}=(a_1^{(1)},a_2^{(1)},a_3^{(1)})$ etc, with $||a^{(1)}||^2=a^{(1)} \cdot a^{(1)}=1$. The $||\cdot ||$ is the Euclidean norm. Similarly, looking at A, the $2$ is associated to $a^{(2)}$. Moreover, for B we have $b^{(1)}$ and $b^{(2)}$. Below, a numerical example of $S \leq 2$ violating setting combinations will be given.

Before entering into more details, the author would like note that we can look upon a CHSH experiment as the question if  measure theoretic {\it or} quantum probability is ruling entanglement. 

\subsection{Correlation in experiment}
Here we answer the question how to obtain in experiment the $E$ values to be used in (\ref{2}). It is technically still impossible to measure directly the $E(a,b)$ for a single pair. The correlation is therefore derived from counting measurement results.  The results enter the raw product moment correlation \cite{GadverNogesGill} to approximate the correlation $E(a,b)$. This is an "averaged over many pairs" correlation. Again, the naive student could respond like: wait a minute, things in quantum theory are already not always what they look like, so how do you know that one pair correlation can be compared to the next and be averaged in experiment? We don't, but we do it anyway.

Suppose we measure $N$ spin pairs. After the last measurement in the series, the correlation $E(a,b)$ is in the experiment of \cite{Weihs} computed approximately. Using the Kronecker delta $\delta_{s,r}$, we count the number of times $S_{A,n}(a_n)=S_{B,n}(b_n)$, and the number of times $S_{A,n}(a_n)=-S_{B,n}(b_n)$ i.e., 
\begin{linenomath*}\begin{equation}
N(=\,|\,a,b)=\sum_{n=1}^N \delta_{S_{A,n}(a), S_{B,n}(b)}\delta_{a_n, a}\delta_{b_n,b},
\end{equation}\end{linenomath*}
and
\begin{linenomath*}\begin{equation}
N(\neq\,|\,a,b)=\sum_{n=1}^N \delta_{S_{A,n}(a), -S_{B,n}(b)}\delta_{a_n, a}\delta_{b_n,b}.
\end{equation}\end{linenomath*}
Hence, we obtain the expression of the correlation
\begin{linenomath*}\begin{equation}\label{4}
E(a,b)=\frac{N(=\,|\,a,b)-N(\neq\,|\,a,b)}{N(=\,|\,a,b)+N(\neq\,|\,a,b)}
\end{equation}\end{linenomath*}
This type of computation of $E$  is also employed in the algorithm and its presented proof of concept in the appendix.

It must be noted that if the researcher employs the inequality $S\leq2$, defined in (\ref{2}), to see in experiment if $E$ in (\ref{4}) gives $S\leq 2$, then implicitly, Bells definition of correlation (\ref{1}) is employed. Hence, a measure theoretic probability is tested in experiment. It is one where $E(a,b)=a\cdot b$ is considered impossible by definition. However, see \cite{HanRecent}.

\section{Preliminaries in the computer design}
 Peres \cite{Peres} formulates it thus: "......, a hidden variable theory which would predict individual
events must violate the canons of special relativity....". Furthermore, the program must mimic an important experiment in the test of locality performed by Weihs  \cite{Weihs}. Note that Weihs's experiment is related to work of Aspect \cite{Aspect}. In Weihs's experiment strict locality conditions were closely approximated and a violation $S>2$ was observed for violating setting combinations of $a$ and $b$ with a quantum correlation $a\cdot b$. 

In \cite{Ge}, however, the present author already showed that there is a nonzero probability that a local hidden variables model may violate the CHSH. Objections to the probability loophole claim in \cite{Ge} were raised in \cite{GadverGil} but were answered in \cite{Geurdes}. The main point is that the employed probability density remains fixed during the trials.
The matter of a possible defective Bell formula was further developed in \cite{HanRecent}. 

It must be noted that the author of \cite{GadverGil} acts as though a random model occurs in \cite{Ge}. However, if $p_1$  and $p_2$ are random numbers between zero and one and $r=p_1+p_2$, then the {\it model} to compute $r$ is fixed, i.e a $+$ operation, despite the fact that the inputs $p_1$ and $p_2$ are random and the outcome $r$ is therefore also random.  The present paper completes the rejection of what has been claimed in \cite{GadverGil} and observes the non-theatrical requirements of \cite{GadverNogesGill}. 
\subsection{Settings}
On the A side Alice has $1\equiv\frac{1}{\sqrt{2}}(1,0,1)$ and $ 2\equiv(\frac{-1}{2},\frac{1}{\sqrt{2}},\frac{1}{2})$ at her disposal. On the B side, Bob has $1\equiv(1,0,0)$ and $2\equiv(0,0,-1)$. For the ease of the argument we inspect, $E(a,b)=a\cdot b$. A simple computation then shows that for a quantum outcome we would see $E(1,1)=1/\sqrt{2}$, $E(1,2)=-1/\sqrt{2}$ while $E(2,1)=-1/2$ and $E(2,2)=-1/2$. Hence, looking at (\ref{2}), for a quantum value, $S=1+\sqrt{2}>2$ is expected in an experiment. The setting parameters $a$ and $b$ are given a value when the A- and B-wing particles leave the source. In flight we allow B (Bob) to change his setting. 
\subsection{Information hiding}
We note that information hiding between Alice and Bob is the algorithmic realization of strict locality. Furthermore, in the computer simulation, A doesn't know anything about B and vice versa. 
All computations are "encapsulated" i.e. local, despite the fact that in the proof of concept (POC), they occur in a single loop (viz. the appendix). In the POC both the A section (Alice) and the B section (Bob) make use of the produced discrete variables created in the source section. This is the computational equivalent of S sending entangled particles to A and B. 

\subsection{Notation}
In the formalism of the algorithm presented below there are {\it no} measures in the sense of measure theoretical distributions. 
We are dealing with arrays of variables, variables as entries of those arrays and functions of those variables. 
Most of the variables and functions project into $\{-1,1\}$. The setting array variables project into $\{1,2\}$. Index variables, most of the time denoted with e.g. $n,m,k$, are integer positive numbers, i.e. $n\in \{0,1,....N\}$, with $N\in\mathbb{N} (N\geq1)$.
\section{Design of the algorithm based on a local model  }
\subsection{Random sources}
In the first place let us introduce random sources to represent random selection of setting. We look at the randomness from the point of view of creating an algorithm. If there are $N$ trials, i.e particle pairs, in the experiment then e.g. two independent random sources can be seen as two arrays with index running  from $1$ to $N$. If $\mathcal{N}_N=(1,2,3,...,N)$, then we, initially, define three random source arrays
\begin{linenomath*}\begin{eqnarray}\label{Rs1}
\begin{array}{ll}
\underline{\mathcal{R}}_{AS}=\sample(\mathcal{N}_N)\\
\underline{\mathcal{R}}_{B}=\sample(\mathcal{N}_N)\\
\underline{\mathcal{R}}_{C}=\sample(\mathcal{N}_N)
\end{array}
\end{eqnarray}\end{linenomath*}
Technically, the map $\mathcal{N}_N \mapsto \underline{\mathcal{R}}_{\cdot}$ is 1-1 but randomized. 
As an example, suppose we have $\mathcal{N}_5=(2,3,5,1,4)$ and so, $\mathcal{N}_{5,1}=2$. Then in the first trial $n=1$, the $\mathcal{N}_{5,1}$ - th element of another array, e.g. $q=(0.1,0.4,-0.9,1.2,1.0)$ is randomly selected, hence, $q(n=1)=0.4$. In the second trial, looking at $\mathcal{N}_5$, we see, $\mathcal{N}_{5,2}=3$ so $q(n=2)=-0.9$, etcetera. Note that this two array procedure is similar to rolling a five-sided dice. If e.g. $\mathcal{N}_5$ is replaced by $\mathcal{M}_{10}$ and multiples are allowed, such as in e.g. $\mathcal{M}_{10}=(2,3,5,1,4,4,5,1,3,3)$ this $q$ "dice" will in 10 turns show three times the side with $-0.9$.

In this way a random source $\underline{\mathcal{R}}$ can be employed in a program and be looked upon as a physical factor giving rise to randomness. 
In a certain sense it refers to 'tHoofts \cite{Hooft} deterministic law hidden inside "randomness". The "freely tossing of a coin" is now replaced with "freely randomizing" the $\underline{\mathcal{R}}_X $  by filling it with $\sample(\mathcal{N}_N)$. There can be no fundamental objection to this particular two array form of randomizing. 

\subsection{Design time settings}
Experimentalists may claim the construction of their measuring instruments. Hence, servers in the experiment may be tuned in design time. There is no fundamental reason to reject design time to the designer of a computer experiment. There is also no reason in physics theory to reject the denial of access of the observers Alice and Bob to information put in the system by the designer during design time. 

Because there is a flow of particles between the S and the A this sharing, i.e. $\underline{\mathcal{R}}_{A}=\underline{\mathcal{R}}_{S}=\underline{\mathcal{R}}_{AS}$, cannot be prevented at run time in a real experiment.  The latter is reflected in the infrastructure of servers in the numerical experiment. The $\{a_n\}_{n=1}^N$ in the experiment are based on the $\underline{a}$ array and the $\underline{\mathcal{R}}_{A}$. For instance $\underline{a}=(1,2,1,2,1,2,...)$. In design time the designer is allowed to introduce a spin-like variable $\sigma_n \in \{-1,1\}$ in the S computer. In the sequence of trials, the variable $\sigma_n$ is selected from $\underline{\sigma}=(-1,1,-1,1,-1,1,...)$. 

We may note that, in case of $\underline{\mathcal{R}}_{A}=\underline{\mathcal{R}}_{S}$, then {\it because of}  $\underline{\mathcal{R}}_{A}=\underline{\mathcal{R}}_{S}$, the relation $a_n=1+\frac{1}{2}(1+\sigma_n)$ occurs on the A side of the experiment. The setting $a_n$ can be either 1 or 2 and is already presented in terms of selection unit parameter vectors in $\mathbb{R}^3$.

Note that the variable $\sigma_n$ can be send to Bob and to Alice without any additional information conveying its meaning. So, Bob cannot derive anything from $\sigma_n$ even though the designer knows the relation. This is because Bob is only active in run time, not in design time.

Finally, the source may also send a $\zeta_n \in \{-1,1\}$ to both Alice and Bob. The $\zeta_n$ in the experiment is based on the  $\underline{\mathcal{R}}_{C}=\sample(\mathcal{N}_N)$ and derives from a $\underline{\zeta}$ array. 

The second random source, $\underline{\mathcal{R}}_{B}$ is used by B exclusively, the third random source, $\underline{\mathcal{R}}_{C}$ is used by the source exclusively.  There appears to be no physical arguments why the sketched configuration is a violation of locality or cannot be found in nature. 

\subsection{Random sources $\mathcal{R}_{\cdot}$ and particles}
The source sends a $\sigma_n \in \{-1,1\}$ and a $\zeta_n \in \{-1,1\}$. to both A and B. In a formal format,
\begin{linenomath*}\[
[A(a_n)]\leftarrow(\sigma,\zeta)_n\leftarrow [S] \rightarrow (\sigma,\zeta)_n\rightarrow [B(b_n)]
\]\end{linenomath*}
Here, e.g. $[A(a)]$ represents the measuring instrument A where Alice has the $a$ setting. This setting "runs synchronous" with $\sigma$ in the source because of the "shared" random source. The particle pair source is represented by $[S]$.

The $\sigma$ and $\zeta$ going into the direction of A are equal to the $\sigma$ and $\zeta$ going to B. Each particle is, in the algorithm, a pair $(\sigma,\zeta)_n$.

\subsection{A side processing of the ($\sigma,\zeta)_n$}
Firstly, let us for the ease of the presentation define a $\sigma_{A,n}=\frac{1+\sigma_n}{2}$. The $\sigma_n$ at the n-th trial from the source $S$ is a result of the sharing of $\underline{\mathcal{R}}_{AS}$. The way the information is used remains hidden to B in order to maintain locality in the model. So, secondly, we have the setting $a_n =\sigma_{A,n}+1$.  Furthermore, we define two functions $\varphi_{A,n}^{-}=\sigma_{A,n}$ and $\varphi_{A,n}^{+}=1-\sigma_{A,n}$. The two functions, together with $\zeta_n$ produce, in turn, a function 
\begin{linenomath*}\[
f_{\zeta_n}(a_n)=\zeta_n\varphi_{A,n}^{+} - \varphi_{A,n}^{-}
\]\end{linenomath*}
Note that $f_{\zeta_n} \in \{-1,1\}$. Hence, we can store the outcome of the computations on the A side immediately in an $N$-size array $S_{A,n}$, together with $a_n$, for trial number $n$ and $n=1,2,3,....N$.

\subsection{B side processing of the ($\sigma,\zeta)_n$}
In the first place, let us determine with the B associated random source, $\underline{\mathcal{R}}_B$, the setting $b_n$. 
Then, secondly and similar to the case of A, but of course completely hidden from A, the $(\sigma,\zeta)_n$ information from the source is processed. We have, $\sigma_{B,n}=\frac{1+\sigma_n}{2}$, then  $\varphi_{B,n}^{-}=\sigma_{B,n}$ and $ \varphi_{B,n}^{+}=\sigma_{B,n}+(\delta_{1,b}-\delta_{2,b})(1-\sigma_{B,n})$. This leads to the function
\begin{linenomath*}\[
g_{\zeta_n}(b_n)=\zeta_n \varphi_{B,n}^{+}+\frac{1-\zeta_n}{\sqrt{2}}\varphi_{B,n}^{-}
\]\end{linenomath*}
For $g_{\zeta_n}(b_n)$ we may note that it projects in the real interval $[-\sqrt{2},\sqrt{2}]$. If $\sigma_{B,n}=1$ then $g_{\zeta_n}(b_n)=1$ for $\zeta_n=1$ and $\sqrt{2}-1$ for $\zeta_n=-1$. If $\sigma_{B,n}=0$, then $\varphi_{B,n}^{-}=0$ and $g_{\zeta_n}(b_n)=\pm 1$.

Hence, in order to generate a response in $\{-1,1\}$, a random $\lambda_2$ from the real interval $[-\sqrt{2},\sqrt{2}]$ is uniformly drawn and $S_B(b_n)=S_{B,n}=\sgn(g_{\zeta_n}(b_n)-\lambda_2)$ in the $n$-th trial. 
We note that as long as Bob doesn't know the meaning of $\sigma_{B,n}$, derived from $\sigma_n$ and related to the $\underline{\mathcal{R}}_{AS}$, locality is warranted. Bob, like Alice, doesn't have access to the design time information.

\subsection{Computer infrastructure}
In computer infrastructure terms one can imagine cables running from the source server running to the A server and running from S server to the B server. One cable, $\mathcal{C}_{SA}(\sigma_n)$, carries the $\sigma_n$ from S to A and the other cable, $\mathcal{C}_{SB}(\sigma_n)$, carries the copy $\sigma_n$ from S to B. Secondly, a cable, $\mathcal{C}_{SA}(\zeta_n)$, carries the $\zeta_n$ from S to A and a cable $\mathcal{C}_{SB}(\zeta_n)$  carries the copy $\zeta_n$ from S to B. In addition to these four cables, a fifth cable, $\mathcal{C}_{AS}(\underline{\mathcal{R}}_A)$ is only used by A to share the (information of) $\underline{\mathcal{R}}_A$ with S. This cable is open {\it only once} and carries only one "pulse" that informs S about the random source at A. The exact time when $\underline{\mathcal{R}}_{A}$ is shared, is in the program at an $n=0$ trial or particle pair. However, the $n=0$ particle pair is merely used for computational convenience and clarity.
\subsubsection{Pre-determined}
In addition, we note the following. The sharing of $\underline{\mathcal{R}}_A$ can also be accomplished in a way that tHooft \cite{Hooft} would most likely call  pre-deterministic. 
In this case server A and S share a common array of randomly distributed integer numbers in an indefinite large array. 
Think of the large array as a design feature like a shared identical random table between computer A and computer S. 

In design time, the A computer receives its $N$ settings $1$ or $2$. This process starts at time $t_{start}$ and ends at $t_{fin}>t_{start}$. 
In computer A a sub-section of the indefinite large array is identified by $t_{start}$ and  $t_{fin}$. This timing mechanism runs parallel in S where a copy of the large array of A resides. Note that a transformation of the numbers in the sub-section can be performed to reach a similar numerical form as if we would have performed $\underline{\mathcal{R}}_A=$sample$\left(\mathcal{N}_N\right)$. In this case $\underline{\mathcal{R}}_A$ and $\underline{\mathcal{R}}_S$ are different and the "bridging" between A and S is done via the computation of $\underline{\pi}$ in S. Here, $\pi_n=1$ when $\underline{\sigma}_{\left(\underline{\mathcal{R}}_A\right)_n}=\underline{\sigma}_{\left(\underline{\mathcal{R}}_S \right)_n}$ and $\pi_n=-1$ when $\underline{\sigma}_{\left(\underline{\mathcal{R}}_A\right)_n} \neq \underline{\sigma}_{\left(\underline{\mathcal{R}}_S \right)_n}$. In the pre-determined case the $\sigma_n$ is computed like
\[
\sigma_n=\pi_n \underline{\sigma}_{\left(\underline{\mathcal{R}}_S\right)_n}
\]
\subsubsection{Mr. X}
It must also be noted that nature is neutral in the following sense. Two situations of feeding parameter settings into measuring instruments may arise. Firstly, looking at the A side, a proverbial Mr. X is sitting in front of A and, before $(\sigma,\zeta)_n$ enters the measurement area of A, Mr. X has selected with a coin the $a_n$. Secondly we have the case where Mr. X delivers his $N$ coin tosses {\it before } the experiment starts and A runs on a batch input of Mr. X's coin tosses. In both cases Alice only makes a record of the $a_n$ and the resulting $\pm 1$ output. We can imagine a Mr. Y at Bob's side and Bob only recording the outcome.
The difference between on-the-spot hand-fed entrance of coin toss values for $a_n$ or batch processing of previous series is insignificant to the problem and a mere illusion. 
This is so because we may assume that Nature has no eyes to witness the difference between the activities of Mr. X. 
The reason is that the incoming particle  makes contact with the instrument which has a certain setting. The incoming particle does not make contact with Mr. X who is delivering the settings.
Hence, the use of previously selected setting series, "from the days of Hamurabi" \cite{GadverdeGadverWeerDieGil},  that are on-line, just-in-time, and hand-fed into A, can equally well and without any violation of experimental protocol be processed in a batch and translated into $\underline{\mathcal{R}}_{A}$. The $\underline{\mathcal{R}}_{A}$ can subsequently be shared with S. 

It must be noted also that the proverbial Mr. X may freely use {\it one} on-line hand-fed set from Hamurabi's days at the side of B. Mr. Y is then posted at Alice and makes a setting entrance batch possible. Because Mr. X has no knowledge of design time and the labels A and B can be arbitrarily interchanged, the selection of Mr. X with the single on-line hand-fed set from the days of Hamurabi will have a nonzero probability to violate the CHSH with the local algorithm provided in the paper. 
Of course, the next step is the ad hoc requirement to have {\it two} ancient hand-fed inputs. But before doing that, Mr. X must explain if Nature at A would really note the difference. If not, then this requirement only has theatrical value. If yes, then the choice of having a Hamurabi set, is allowed into the design too. Suppose Simon is running the source. Then why would the data from Hamurabi's days be accessible to Mr. X and not to Simon? The setting at Bob remains at all times random.
Again, it is unlikely that Nature in A will behave differently when Mr. X is holding a tablet with cuneiform markings and feeding just-in-time 1's and 2's into the selection area (the little green rectangle in the A rectangle of figure - \ref{Predet}) contrasted with the situation where Mr. Y helps Mr. X with the translation and writes down the 1's and 2's from the tablet first and put them afterwards in the A area to derive a random source $\underline{\mathcal{R}}_A$ etcetera. 
The question is, how far must one allow the incorporation of theatrical requirements in the design. Depending upon the amount of omniscience a stakeholder thinks he or she has, theatrical requirements are thought necessary.

It is an other matter, whether or not Mr. X is able to determine the difference between  e.g. a $\underline{\mathcal{R}}_C=$sample$\left(\mathcal{N}_n\right)$ and $\underline{\zeta}=(-1,1,\dots )$ random process and a coin toss for obtaining $\zeta_n \in \{-1,1\}$, with $n=1,2,\dots, N$. Here a test resembling a Turing test can be invoked \cite{Say}. A computer generates 1's and -1's with random source and value arrays versus a human tosses a coin and generates a series of 1's and -1's. Both processes are covert. Mr. X has to decide if there is a difference. The claim is that Mr. X cannot detect the difference better than chance.
\begin{linenomath*}
\begin{figure}
\caption{Explicit key concepts of computer infrastructure in pre-determined format. The $\underline{\mathcal{R}}_A$ is shared between A and S. In S, a computation of an auxiliary array $\underline{\pi}$ is run at $n=0$ to generate the $\sigma_n$ for $n>0$. 
The green rectangles in the A and B are information inflow areas. The blue rectangles are for information outflow. If e.g. Bob selects a setting in the $n$-th trial ($n>0$) he pushes a button and receives from B a setting either $1$ or $2$. He then puts the setting in the setting area of his computer (the little green top rectangle in the B rectangle). Then the $(\sigma,\zeta)_n$ enters via the big green rectangle in B. The B-side algorithm starts running. After some time a response $\pm1$ comes out of the B computer and Bob makes a record of it together with the setting $b_n$. Similar case for Alice. The red lines hide the S activity from Alice and Bob.}\label{Predet}
\includegraphics[scale=0.50]{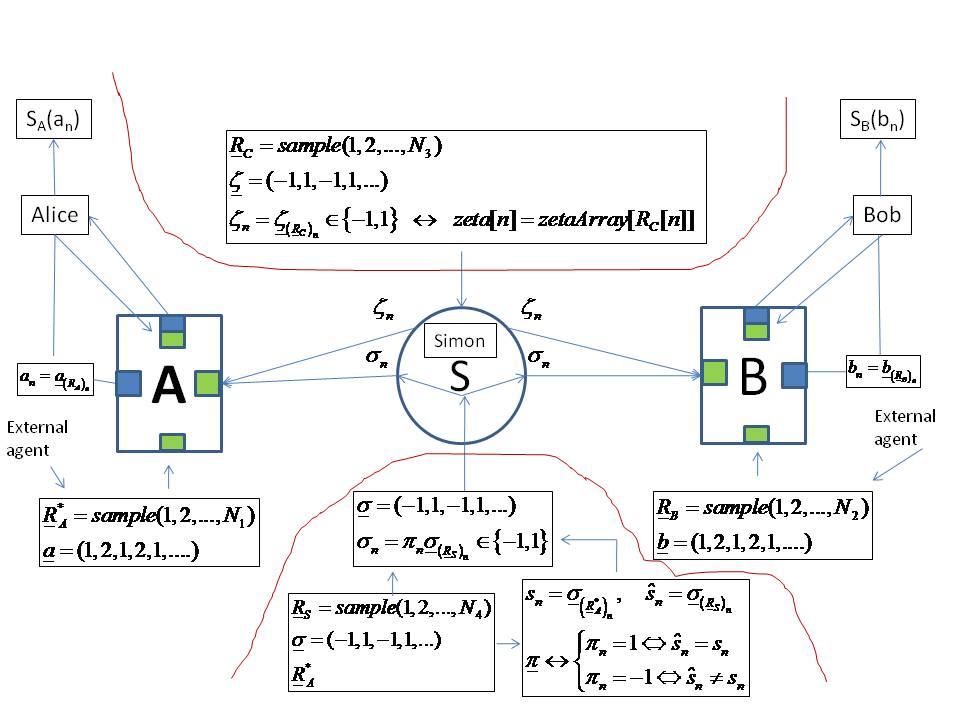}
\end{figure}
\end{linenomath*}
\newpage
\section{Conclusion \& discussion}
In the paper a simple design is given that is able to violate the CHSH inequality with numerical values close to the expected quantum mechanics. The reader kindly notes that no violation of locality is employed. B doesn't know the meaning of the A-S shared information send to B. The information from S to Alice is inaccessible to Bob.
Both Alice and Bob are not allowed access to the design. Further, if the sharing of information runs along the lines of 'tHooft's pre-determinism \cite{Hooft} then S and A do not know they share information, i.e the random source array, $\underline{\mathcal{R}}_A$.

The reader also notes that the computer set-up is designed to explain the outcome of the A-S-B experiment such as in Weihs's \cite{Weihs}  and should not be confused with experimental configurations unequal to $A(a)\leftarrow S \rightarrow B(b)$. 

In the appendix, the essential loops in the R program over $n=1,2,3...N$ are presented. 
The code is the POC of the algorithm with $\underline{\mathcal{R}}_A=\underline{\mathcal{R}}_S=\underline{\mathcal{R}}_{AS}$. 
This situation refers to "pre-trials information leakage" from A to S. In the case of a pre-determined format, such as in figure - \ref{Predet}, we have,  $\underline{\mathcal{R}}_{A}$, $\underline{\mathcal{R}}_{S}$, $\underline{\mathcal{R}}_{C}$ and $\underline{\mathcal{R}}_{B}$. 
This is not an active information leakage but a wired-in sharing of information in design time. In the latter case $\underline{\pi}$ is auxiliary to the computation of the $\sigma_n$. The reader is referred to figure - \ref{Predet}.  It is noted that nobody knows if, either via A-S leakage or via wired-in pre-determined sharing, the measuring instrument, A, and the particle source, S, share information yes or no. The use of encapsulating information and a distinction between design time and runtime also makes sure that A and S are unaware that they share information. 

In both cases we assumed a $n=0$ initial particle pair to do the necessary initial computations. Furthermore, there is a flow of particles between $S$ and $A$. From $S$ to $A$ the flow is "forced" by the experimenter. In this design, flow of information from $A$ to $S$ can be enforced by nature on the experimenter. It is perhaps like 'tHooft claimed: ".... every no-go theorem comes with small print" \cite{Hooft}. 

To this it must be added that the CHSH is based on Bell's formula. In turn, Bells formula is based on the (probabilistic) distribution of hidden variables $\lambda$. The POC computer program is a typical "classical" algorithm. 
The set-up of three computers is a realization of a classical system that mimics the instruments in the experiment. 
How would an opponent of extra local parameters interpret the numerically obtained violation, other than the rejection of the necessity of quantum probabilities to violate the CHSH?

As required by the author of \cite{GadverGil} a computer simulation rejects the criticism raised in \cite{GadverGil}. We may claim this because our "freezing the setting of $a$  at particle creation" is a valid CHSH type of experiment. It would be strange to say that locality and causality cannot occur in an experiment where "in flight" changes in both wings are allowed whereas one must admit that locality and causality occurs when only B wing "in flight" changes of setting may occur. This is all the more so because the A and B role in the simulation can be selected randomly.

The metaphor requirements of \cite{GadverNogesGill} are met or are identified as solvable within the design. Note that  a violation of the CHSH criterion would, most likely, not have been possible without a probability loophole in the CHSH \cite{Ge}. Furthermore, the freely selected settings are created at design time.  
It is moreover, hard to see how a particle pair in a distant source would behave differently when Alice and Bob or an external agent such as e.g. as Mr. X, employ, to them, unknown random sequences  $\underline{\mathcal{R}}_{A}$  and $\underline{\mathcal{R}}_{B}$  for their setting selection, compared to the case where a coin to select the respective setting is employed. 
We note that preformed but just-in-time hand fed ancient setting sequences by Mr. X would give a nonzero probability of CHSH violation with a local algorithm. 
In this sense progress is made when looking at \cite{Ge}. In the latter case only a nonzero number of  quartets of setting values $\{(1,1), (1,2), (2,1), (2,2)\}$, violate the CHSH with local means. 
Finally, if the behavior of Mr. X matters at quantum level, then it must be entered into the design and can be incorporated to solve that challenge in this way. 
The required hardware of the computer experiment is: three computers, four cables, a timing mechanism for sending pulses $(\sigma,\zeta)_n$ from S to A and S to B and a fifth cable from A to S that is used only once in a $n=0$ pre-experimental statistical trial.

Furthermore, testers must be completely unaware of the design time activities of the designer.
The reason is that, obviously, experimental physicists and their assistants were not cognitively present when in the big bang, matter was created.
Implementation of the software on  the A and B side algorithms plus detector timers are required together with the algorithm for S. 
The design of the infrastructure for a predetermined format, which also can be build with the hardware given above, is provided in figure - \ref{Predet}.  
Of course contingency programming for $n=0$ needs to be done such that no single particle pair lacks from counting. 
In tests on an ordinary computer, a maximum of $N=1\times 10^{7}$ number of particle pairs was reached. 

We claim that we are allowed to say that the present result corrects Peres' statement \cite{Peres},  that violations of the CHSH inequality "violate the canons of special relativity".
We also add here that serious doubts can be cast on the mathematical consistency of Bells methodology \cite{HanRecent}. To the present author this mathematical deficit of Bell methodology represents an additional reason to maintain the idea of local hidden variables in the sense of additional parameters to supplement the wave function. 
However, to quote Einstein\cite{Eins} "We believe ... that such a [more complete ?] theory exists" 
. Sure, this appears quite easily said but far more difficult to be obtained.

\newpage
\appendix{Appendix:}
The $\underline{\mathcal{R}}_{A}=\underline{\mathcal{R}}_{S}=\underline{\mathcal{R}}_{AS}$ algorithm is shown in the POC. 
\begin{verbatim}
N<-4e5
a<-array(0,N)
aKeep<-array(0,N)
sigma<-array(0,N)
zeta<-array(0,N)
b<-array(0,N)
bKeep<-array(0,N)
RAS<-sample(seq(1,N),N,replace=FALSE,prob=NULL)
RB<-sample(seq(1,N),N,replace=FALSE,prob=NULL)
RC<-sample(seq(1,N),N,replace=FALSE,prob=NULL)
#
for(j in 1:N){
  k<-as.integer(j/2)
  m<-j/2
  if(m==k){
    a[j]<-2
    b[j]<-2
    sigma[j]<-1
    zeta[j]<-1
  }else{
    a[j]<-1
    b[j]<-1
    sigma[j]<-(-1)
    zeta[j]<-(-1)
  }
}
#
scoreA<-array(0,c(2,N))
scoreB<-array(0,c(2,N))
for (n in 1:N){
#Source section  
  zetah<-zeta[RC[n]]
  sygma<-sigma[RAS[n]]
#A section
  aSet<-a[RAS[n]]
  aKeep[n]<-aSet
  phiAmin<-((sygma+1)/2)
  phiAplus<-1-((sygma+1)/2)
  f<-zetah*phiAplus-phiAmin
  scoreA[aSet,n]<-f
#B section
  phiBmin<-((sygma+1)/2)
  bSet<-b[RB[n]]
  bKeep[n]<-bSet
  if(((sygma+1)/2)==1){
    phiBplus<-1
  }else{
    if(bSet==1){
      phiBplus<-1
    }
    if(bSet==2){
      phiBplus<-(-1)
    }
  }  
  g<-zetah*phiBplus 
  g<-g+((1-zetah)*phiBmin/sqrt(2))
  lambda_2<-runif(1)*sqrt(2)
  lambda_2<-sign(0.5 - runif(1))*lambda_2
  scoreB[bSet,n]<-sign(g-lambda_2)
}
E<-matrix(0,nrow=2,ncol=2)
Neq<-array(0,c(2,2))
Nneq<-array(0,c(2,2))
for (n in 1:N){
  aSet<-aKeep[n]
  bSet<-bKeep[n]
  if (scoreA[aSet,n]==scoreB[bSet,n]){
    Neq[aSet,bSet]<-Neq[aSet,bSet]+1
  }else{
    Nneq[aSet,bSet]<-Nneq[aSet,bSet]+1
  }
}
for(aSet in 1:2){
  for(bSet in 1:2){
    E[aSet,bSet]<-(Neq[aSet,bSet]-Nneq[aSet,bSet])/(Neq[aSet,bSet]+Nneq[aSet,bSet])
  }
}
print(N)
print(E)
CHSH<-E[1,1]-E[1,2]-E[2,1]-E[2,2]
print(paste0("CHSH=",CHSH))

\end{verbatim}


\begin{thebibliography}{}
\bibitem{Jammer} \textsc{M. Jammer}, \textit{The philosophy of quantum mechanics} (Wiley-Interscience, New York, 1974).

\bibitem{Merz}  \textsc{E. Merzbacher}, \textit{Quantum Mechanics} (Wiley \& Sons, New York, London, 1970).

\bibitem{Hameka} \textsc{H.F. Hameka}, \textit{Quantum Mechanics, a conceptual approach}, (Wiley-Interscience, New Jersy, 2004).

\bibitem{Rae} \textsc{A. Rae}, \textit{Quantum Mechanics}, (IOP Publishing, London, 2002).

\bibitem{Grein} \textsc{W. Greiner}, \textit{Relativistic quantum mechanics, Wave equations}, (Third edition, Springer, Berlin, Heidelber, 2000).

\bibitem{Peres} \textsc{A. Peres}, \textit{Quantum theory: Concepts and Methods} (Kluwer Academic Publ., New York, 2002) 165-172.

\bibitem{Haag} \textsc{R. Haag}, \textit{Trying to divide the universe} in A. Borowiec, W. Cegla, B. Jancewicz, \& W. Karowski, Theoretical physics fin de sciecle, (Springer-Verlag, Berlin, Heidelberg 2000)

\bibitem{Eins}
\textsc{A. Einstein, B. Podolsky, N. Rosen,} (1935),
\textit{Can quantum-mechanical description of physical reality be considered complete, Phys. Rev.} \textbf{47} 777-780.

\bibitem{Bell}
\textsc{J.S. Bell, } (1964),
\textit{On the Einstein Podolsky Rosen paradox, Physics} \textbf{1} 195-200.

\bibitem{Clauser}
\textsc{J.F. Clauser,  M.A. Horne, A. Shimony,   R.A. Holt,} (1969),
\textit{Proposed experiment to test local hidden-variables theories, Phys. Rev. Lett.} \textbf{23} 880-884.

\bibitem{Bohm:1983}\textsc{
J. A. Wheeler, and W. H. Zurek,} \textit{Quantum theory and measurement}  (Princeton Univ Press, Princeton, 1983) 356-369.

\bibitem{Bohm2} \textsc{D. Bohm},  \textit{Quantum theory} (Prentice-Hall, New York), 614.

\bibitem{Weihs}
\textsc{G. Weihs, T. Jennewein, C. Simon, H. Weinfurter \& A. Zeilinger,} (1998), 
\textit{Violation of Bell's inequality under strict Einstein locality conditions, Phys. Rev. Lett.} \textbf{81}, 5039-5043.

\bibitem{Aspect}
\textsc{A. Aspect}, (1976), \textit{Proposed experiment to test the nonseparability of quantum mechanics. } ,
\textit{Phys. Rev.} \textbf{D14}, 1944-1955. 

\bibitem{Ge}
\textsc{J.F. Geurdes,} (2014),
\textit{A probability loophole in the CHSH, Results in Physics} \textbf{4}, 81-82.

\bibitem{Hooft} \textsc{Gerard 't Hooft}, (2001), How does God play dice ?, arxiv.org/abs/hep-th/0104219v1, 3.

\bibitem{GadverGil}
\textsc{R.D. Gill,} (2015), 
\textit{No probability loophole in the CHSHS, Results in Physics} \textbf{5}, 156-157.
http://dx.doi.org/10.1016/j.rinp.2015.06.002.

\bibitem{GadverNogesGill}
\textsc{R.D. Gill,} (2003), Time, Finite Statistics, and Bell’s Fifth Position, arxiv.org/abs/quant-ph/0301059v1.

\bibitem{GadverdeGadverWeerDieGil} \textsc{R.D. Gill,} Remark made in a conversation (2006).

\bibitem{Geurdes}
\textsc{J.F. Geurdes,} (2015), 
\textit{Why one can maintain that there is a probability loophole in the CHSH} arXiv:1508.04798.

\bibitem{Say}\textsc{A.P, Saygin, G. Roberts, \&  G. Beber}, \textit{Comments on "Computing machinery and Intelligence" by Alan Turing}, in R Epstein, G. Roberts, G. Poland, Parsing the Turing test (Springer, Dordrecht, 2008, doi:10.1007/978-1-4020-6710-5.)

\bibitem{HanRecent}
\textsc{J.F. Geurdes, K. Nagata, T. Nakamura and A. Farouk } (2017), 
\textit{A note on the possibility of incomplete theory} arXiv:1704.00005.
\end{thebibliography}
\end{document}